\documentclass [11pt,a4paper] {article}

\usepackage[cp1252]{inputenc}

\usepackage{amssymb}
\usepackage{amsmath}
\usepackage{amsfonts,amssymb}
\usepackage[dvips]{graphicx}
\usepackage{bbm}
\usepackage{enumerate}

\usepackage{amsthm}

\usepackage{cancel}

\DeclareMathAlphabet{\mathpzc}{OT1}{pzc}{m}{it}

\begin{document}

\title{Do quantum propositions obey the principle of excluded middle?}

\author{Arkady Bolotin\footnote{$Email: arkadyv@bgu.ac.il$\vspace{5pt}} \\ \textit{Ben-Gurion University of the Negev, Beersheba (Israel)}}

\maketitle

\begin{abstract}\noindent The present paper demonstrates the failure of the principle of excluded middle (PEM) in the lattice of all closed linear subspaces of a Hilbert space (that is usually defined as quantum logic). Namely, it is shown that for a qubit, a proposition and its negation can be both false. Since PEM is the assumed theorem of quantum logic, this raises the question: If PEM holds in the orthocomplemented lattice of all propositions of the quantum system, then how the failure of PEM in quantum logic can be explained? Alternatively, if the propositions relating to the quantum system do not obey PEM, then what is the semantics of those propositions? Possible answers to these questions are analyzed in the present paper.\\

\end{abstract}

\section{Introduction and preliminaries}  

\noindent Recall that \textit{a qub\/it} is a two-state quantum-mechanical system. Correspondingly, any pure qubit state $|\Psi_n^{(Q)}\rangle \notin \{0\}$ (where $\{0\}$ is the zero-vector space containing only vector $0$) can be represented as a linear superposition of two states $|\Psi_1^{(R)}\rangle \notin \{0\}$ and $|\Psi_2^{(R)}\rangle \notin \{0\}$ (such that $|\Psi_n^{(Q)}\rangle \neq |\Psi_1^{(R)}\rangle,\, |\Psi_2^{(R)}\rangle$), namely,\smallskip

\begin{equation} \label{SUP} 
   |\Psi_n^{(Q)}\rangle
   =
   c_1 |\Psi_1^{(R)}\rangle + c_2 |\Psi_2^{(R)}\rangle
   \;\;\;\;  ,
\end{equation}
\smallskip

\noindent where $n \in \{1,2\}$, $Q,R \in \{1,2,3\}$, and $c_1,c_2 \in \mathbb{C}$. The states $|\Psi_n^{(Q)}\rangle$, $|\Psi_1^{(R)}\rangle$ and $|\Psi_2^{(R)}\rangle$ are the eigenvectors of the projection operators $\hat{P}_n^{(Q)}$, $\hat{P}_1^{(R)}$, and $\hat{P}_2^{(R)}$, respectively, defined by the common formula through the Kronecker delta $\delta_{ab}$ calculation, explicitly,\smallskip

\begin{equation} \label{KRON} 
   \hat{P}_a^{(b)}
   =
   \frac{1}{2}
      \left[
         \begin{array}{l l}
            1-(-1)^a \delta_{b3}
            &
            (-1)^a (-\delta_{b1}+i\delta_{b2})
            \\
            (-1)^a (-\delta_{b1}-i\delta_{b2})
            &
            1+(-1)^a \delta_{b3}
         \end{array}
      \right]
   \;\;\;\;  .
\end{equation}
\smallskip

\noindent \textit{The column spaces} (a.k.a. \textit{images} and \textit{ranges}) of the operators $\hat{P}_n^{(Q)}$ denoted as $\mathrm{ran}(\hat{P}_n^{(Q)})$ are closed linear subspaces of the Hilbert space $\mathcal{H}$. Explicitly, $\mathrm{ran}(\hat{P}_n^{(Q)})$ is the subset of the vectors $|\Psi\rangle \in \mathcal{H}$ that are in the image of $\hat{P}_n^{(Q)}$, i.e.\smallskip

\begin{equation} 
   \mathrm{ran}(\hat{P}_n^{(Q)})
   \equiv
   \left\{
      |\Psi\rangle \in \mathcal{H}
      : \;
      \hat{P}_n^{(Q)}|\Psi\rangle = |\Psi\rangle
   \right\}
   \;\;\;\;  .
\end{equation}
\smallskip

\noindent Dually, $\mathrm{ran}(\hat{1} - \hat{P}_n^{(Q)}) = \mathrm{ker}(\hat{P}_n^{(Q)})$ stands for \textit{the null space} (a.k.a. \textit{kerne\/l}) of the projector $\hat{P}_n^{(Q)}$, i.e., the subset of the vectors $|\Psi\rangle \in \mathcal{H}$ that are mapped to zero by $\hat{P}_n^{(Q)}$, namely,\smallskip

\begin{equation} 
   \mathrm{ran}(\hat{1} - \hat{P}_n^{(Q)})
   =
   \mathrm{ker}(\hat{P}_n^{(Q)})
   \equiv
   \left\{
      |\Psi\rangle \in \mathcal{H}
      : \;
      \left(\hat{1} - \hat{P}_n^{(Q)}\right)|\Psi\rangle = |\Psi\rangle
   \right\}
   \;\;\;\;  ,
\end{equation}
\smallskip

\noindent where $\hat{1}$ denotes the identity operator. For that reason, the projector $\hat{1} - \hat{P}_n^{(Q)}$ can be understood as the negation of $\hat{P}_n^{(Q)}$, i.e.,\smallskip

\begin{equation} 
   \neg\hat{P}_n^{(Q)}
   =
   \hat{1} - \hat{P}_n^{(Q)}
   \;\;\;\;  .
\end{equation}
\smallskip

\noindent It results from the formula (\ref{KRON}) that\smallskip

\begin{equation} 
   \hat{P}_1^{(Q)} + \hat{P}_2^{(Q)}
   =
   \hat{1}
   \;\;\;\;   
\end{equation}
\smallskip

\noindent for any $Q$. Hence, in the two-dimensional Hilbert space $\mathcal{H}=\mathbb{C}^2$ one has\smallskip

\begin{equation} 
   \neg\hat{P}_n^{(Q)}
   =
   \hat{P}_{n+(-1)^{n-1}}^{(Q)}
   \;\;\;\;  .
\end{equation}
\smallskip

\noindent Consistent with the assumption of Birkhoff and von Neumann \cite{Birkhoff}, the set of all the closed linear subspaces of $\mathbb{C}^2$, namely,\smallskip

\begin{equation}  
   \mathcal{L}(\mathbb{C}^2)
   =
   \left\{
      \mathrm{ran}(\hat{0})
      ,\,
      \mathrm{ran}(\hat{P}_{1}^{(1)\!})
      ,\,
      \mathrm{ran}(\hat{P}_{2}^{(1)\!})
      ,\,
      \dots
      ,\,
      \mathrm{ran}(\hat{P}_{1}^{(3)\!})
      ,\,
      \mathrm{ran}(\hat{P}_{2}^{(3)\!})
      ,\,
      \mathrm{ran}(\hat{1})\!
   \right\}
   \;\;\;\;  ,
\end{equation}
\smallskip

\noindent where $\mathrm{ran}(\hat{0}) = \{0\}$ and $\mathrm{ran}(\hat{1}) = \mathbb{C}^2$, form a complete lattice called the Hilbert lattice $(\mathcal{L}(\mathbb{C}^2),\le)$ where the symbol $\le$ denotes the partial ordering on $\mathcal{L}(\mathbb{C}^2)$. This partial ordering is defined by\smallskip

\begin{equation}  
   \mathrm{ran}(\hat{0})
   \le
   \mathrm{ran}(\hat{P}_n^{(Q)})
   \iff
   \{0\}
   \subseteq
   \mathrm{ran}(\hat{P}_n^{(Q)})
   \;\;\;\;  ,
\end{equation}

\begin{equation}  
   \mathrm{ran}(\hat{P}_n^{(Q)})
   \le
   \mathrm{ran}(\hat{1})
   \iff
   \mathrm{ran}(\hat{P}_n^{(Q)})
   \subseteq
   \mathbb{C}^2
   \;\;\;\;  .
\end{equation}
\smallskip

\noindent Because $(\mathcal{L}(\mathbb{C}^2),\le)$ is complete, it has \textit{join} and \textit{meet} operations denoted $\vee$ and $\wedge$ \cite{Davey, Burris}. Particularly, for each pair of elements $\mathrm{ran}(\hat{P}_n^{(Q)})$ and $\mathrm{ran}(\hat{P}_m^{(R)})$ of $\mathcal{L}(\mathbb{C}^2)$, where $\mathrm{ran}(\hat{P}_n^{(Q)}) \neq \mathrm{ran}(\hat{P}_m^{(R)})$ and $m \in \{1,2\}$, one has\smallskip

\begin{equation}  
   \mathrm{ran}(\hat{P}_n^{(Q)})
   \wedge
   \mathrm{ran}(\hat{P}_m^{(R)})
   =
   \mathrm{ran}(\hat{P}_n^{(Q)})
   \cap
   \mathrm{ran}(\hat{P}_m^{(R)})
    =
   \{0\}
   \;\;\;\;  ,
\end{equation}
\vspace*{-10mm}

\begin{equation}  
   \mathrm{ran}(\hat{P}_n^{(Q)})
   \vee
   \mathrm{ran}(\hat{P}_m^{(R)})
   =
   \left(
      \mathrm{ran}(\hat{P}_{n+(-1)^{n-1}}^{(Q)})
      \cap
      \mathrm{ran}(\hat{P}_{m+(-1)^{m-1}}^{(R)})
   \right)^{\perp}
   =
   \left(\{0\}\right)^{\perp}
   =
   \mathbb{C}^2
   \;\;\;\;  ,
\end{equation}
\smallskip

\noindent where $(\cdot)^{\perp}$ stands for the orthogonal complement of $(\cdot)$.\\

\noindent Let the qubit be prepared in the pure state given by the vector $|\Psi\rangle$ residing in the closed linear subspace $\mathcal{H}_p  \in \mathcal{L}(\mathbb{C}^2)$. Then, the logical proposition $P_n^{(Q)}$ asserting that \textit{this vector l\/ies in the range of the proje\/ction operator $\hat{P}_n^{(Q)}$ on $\mathbb{C}^2$} can be set forth by\smallskip

\begin{equation}  
   P_n^{(Q)}
   \equiv
   \mathrm{Prop}
   \left(
      |\Psi\rangle
      \in
      \mathcal{H}_p
      \wedge
      \mathrm{ran}(\hat{P}_n^{(Q)})
   \right)
   \;\;\;\;  ,
\end{equation}
\smallskip

\noindent where $\mathcal{H}_p \,\wedge \mathrm{ran}(\hat{P}_n^{(Q)})$ stands for the meet operation on the closed linear subspaces $\mathcal{H}_p$  and $\mathrm{ran}(\hat{P}_n^{(Q)})$ of $\mathcal{L}(\mathbb{C}^2)$.\\

\noindent In a dual manner, the proposition asserting that \textit{this ve\/ctor lies in the kernel of the proje\/ction operator $\hat{P}_n^{(Q)}$} can be defined as the negation $\neg P_n^{(Q)}$ of the proposition $P_n^{(Q)}$ and set forth by\smallskip

\begin{equation}  
   \neg P_n^{(Q)}
   \equiv
   \mathrm{Prop}
   \left(
      |\Psi\rangle
      \in
      \mathcal{H}_p
      \wedge
      \mathrm{ker}(\hat{P}_n^{(Q)})
   \right)
   =
   \mathrm{Prop}
   \left(
      |\Psi\rangle
      \in
      \mathcal{H}_p
      \wedge
      \mathrm{ran}(\hat{P}_{n+(-1)^{n-1}}^{(Q)})
   \right)
   \;\;\;\;  ,
\end{equation}
\smallskip

\noindent where $\mathcal{H}_p \wedge \mathrm{ran}(\hat{P}_{n+(-1)^{n-1}}^{(Q)})$ is the meet of $\mathcal{H}_{|\Psi\rangle}$  and $\mathrm{ran}(\hat{P}_{n+(-1)^{n-1}}^{(Q)})$ of $\mathcal{L}(\mathbb{C}^2)$.\\

\noindent In line with these definitions, if the qubit is prepared in the state $|\Psi_m^{(R)}\rangle \notin \{0\}$ belonging to either the range $\mathrm{ran}(\hat{P}_1^{(R)})$ or the range $\mathrm{ran}(\hat{P}_2^{(R)})$, the proposition $P_1^{(R)}$ and its negation $\neg P_1^{(R)}$ are\smallskip

\begin{equation}  
   P_1^{(R)}\!
   =
   \mathrm{Prop}\!
   \left(
      |\Psi_m^{(R)}\rangle
      \in
      \mathrm{ran}(\hat{P}_m^{(R)})
      \!\wedge
      \mathrm{ran}(\hat{P}_1^{(R)})
   \right)\!
   =\!
   \left\{\!
      \begin{array}{l}
         \mathrm{Prop}\left(|\Psi_1^{(R)}\rangle \in \mathrm{ran}(\hat{P}_1^{(R)})\right), \;  m = 1\\
         \mathrm{Prop}\left(|\Psi_2^{(R)}\rangle \in \{0\}\right),                         \;\;\;\,\qquad  m = 2
      \end{array}
   \right.
   \;\;\;  ,
\end{equation}
\vspace*{-5.5mm}

\begin{equation}  
   \neg P_1^{(R)}\!
   =
   \mathrm{Prop}\!
   \left(
      |\Psi_m^{(R)}\rangle
      \in
      \mathrm{ran}(\hat{P}_m^{(R)})
      \!\wedge
      \mathrm{ran}(\hat{P}_2^{(R)})
   \right)\!
   =\!
   \left\{\!
      \begin{array}{l}
         \mathrm{Prop}\left(|\Psi_1^{(R)}\rangle \in \{0\}\right),                         \;\;\;\,\qquad  m = 1\\
         \mathrm{Prop}\left(|\Psi_2^{(R)}\rangle \in \mathrm{ran}(\hat{P}_2^{(R)})\right), \;  m = 2
      \end{array}
   \right.
   \;\;\;  .
\end{equation}
\smallskip

\noindent As $|\Psi_m^{(R)}\rangle \notin \{0\}$, the proposition $\mathrm{Prop}\big(|\Psi_m^{(R)}\rangle \in \{0\} \big)$ must be a contradiction, which means that $P_1^{(R)}$ and $\neg P_1^{(R)}$ \textit{cannot be false to\/ge\/ther}.\\

\noindent However, if the qubit is prepared in the superposition of the states $|\Psi_1^{(R)}\rangle$ and $|\Psi_2^{(R)}\rangle$ shown in (\ref{SUP}), i.e., in the state $|\Psi_n^{(Q)}\rangle \notin \{0\}$ located in the subspace $\mathrm{ran}(\hat{P}_n^{(Q)})$, then the proposition $P_1^{(R)}$ and its negation $\neg P_1^{(R)}$ \textit{are b\/oth false}:\smallskip

\begin{equation} \label{NEG1} 
   P_1^{(R)}\!
   =
   \mathrm{Prop}\!
   \left(
      |\Psi_n^{(Q)}\rangle
      \in
      \mathrm{ran}(\hat{P}_n^{(Q)})
      \!\wedge
      \mathrm{ran}(\hat{P}_1^{(R)})
   \right)\!
   =\!
   \mathrm{Prop}\!
   \left(
      |\Psi_n^{(Q)}\rangle \in \{0\}
   \right)
   \;\;\;  ,
\end{equation}

\begin{equation} \label{NEG2} 
   \neg P_1^{(R)}\!
   =
   \mathrm{Prop}\!
   \left(
      |\Psi_n^{(Q)}\rangle
      \in
      \mathrm{ran}(\hat{P}_n^{(Q)})
      \!\wedge
      \mathrm{ran}(\hat{P}_2^{(R)})
   \right)\!
   =\!
   \mathrm{Prop}\!
   \left(
      |\Psi_n^{(Q)}\rangle \in \{0\}
   \right)
   \;\;\;  .
\end{equation}
\smallskip

\noindent This is confusing because \textit{the princip\/le of exclude\/d middle} (PEM for short), which states that a proposition and its negation cannot be false together \cite{Beziau}, is supposed to hold in the lattice $(\mathcal{L}(\mathcal{H}),\le)$ – i.e., the lattice of all the closed subspaces of a Hilbert space $\mathcal{H}$ – usually defined as \textit{quantum logic} \cite{Redei,Ptak}.\\

\noindent Assuming after \cite{Mackey} that there is an isomorphism between the lattice of quantum propositions $\mathcal{Q}$ – i.e., the orthocomplemented lattice of all propositions of a quantum mechanical system – and the lattice $(\mathcal{L}(\mathcal{H}),\le)$, this raises the question: If PEM holds in $\mathcal{Q}$, then how the failure of PEM in $(\mathcal{L}(\mathcal{H}),\le)$ demonstrated in (\ref{NEG1}) and (\ref{NEG2}) can be explained?\\

\noindent Alternatively, if quantum propositions do not obey PEM, then what is the semantics of those propositions?\\

\noindent The present paper analyzes possible answers to these questions.\\

\section{Intuitionistic quantum logic}  

\noindent Let us start with the analysis of the statement that quantum propositions reject PEM.\\

\noindent From a mathematical point of view, to interpret quantum propositions in terms of the intuitionistic propositional logic (in which PEM is not valid), one must find mathematical objects in quantum theory that form the structure of a Heyting algebra, i.e., a semantic of intuitionistic propositional logic.\\

\noindent Recall that \textit{the pseudo-complement $\neg a$} of an element $a$ of the Heyting algebra $H$ is \textit{the supremum} of the set $\{b:\; b \wedge a = 0\}$ such that $b \in H$ and $a$ belongs to this set, i.e., $a \wedge \neg a = 0$ holds \cite{Yankov}.\\

\noindent Suppose that the elements of the partially ordered set $\mathcal{L}(\mathbb{C}^2)$ are also elements of the Heyting algebra $H$, i.e., the bounded lattice with join and meet operations $\vee$ and $\wedge$ and with least element $\{0\}$ (denoted by 0) and greatest element $\mathbb{C}^2$ (denoted by 1).\\

\noindent Consider such a subset $\mathcal{S}$ of the poset $\mathcal{L}(\mathbb{C}^2)$ that\smallskip

\begin{equation}  
   \mathcal{S}\!
   =
   \mathcal{L}(\mathbb{C}^2)
   \setminus
   \left\{
      \{0\}, \mathbb{C}^2
   \right\}
   \;\;\;\;  .
\end{equation}
\smallskip

\noindent The subset $\mathcal{S}$ has a single upper bound which is greatest element $\mathbb{C}^2$. To be sure, for any pair of ranges $\mathrm{ran}(\hat{P}_n^{(Q)})$ and $\mathrm{ran}(\hat{P}_m^{(R)})$ in $\mathcal{S}$, where $\mathrm{ran}(\hat{P}_n^{(Q)}) \neq \mathrm{ran}(\hat{P}_m^{(R)})$, one has $\mathbb{C}^2 \supseteq \mathrm{ran}(\hat{P}_n^{(Q)})$ and $\mathbb{C}^2 \supseteq \mathrm{ran}(\hat{P}_m^{(R)})$, at the same time as $\mathrm{ran}(\hat{P}_n^{(Q)})$ and $\mathrm{ran}(\hat{P}_m^{(R)})$ are \textit{incomp\/ara\/b\/le} with each other. This means that any set\smallskip

\begin{equation}  
   \left\{
      \mathrm{ran}(\hat{P}_n^{(Q)})
      :
      \;\;
      \mathrm{ran}(\hat{P}_n^{(Q)}) \wedge \mathrm{ran}(\hat{P}_m^{(R)})
      =
      \{0\}
   \right\}
   \;\;\;\;   
\end{equation}
\smallskip

\noindent does not contain a minimal element, i.e., a pseudo-complement. Hence, the closed linear subspaces of the Hilbert space $\mathbb{C}^2$ cannot form a pseudo-complemented lattice and, as a result, an acceptable Heyting algebra $H$.\\

\noindent Still, there is a possibility to replace the Hilbert lattice $(\mathcal{L}(\mathbb{C}^2),\le)$ with some other (distributive) lattice that defines a Heyting algebra.\\

\noindent For example, in the paper \cite{Caspers}, as a replacement of $(\mathcal{L}(\mathbb{C}^2),\le)$, the lattice $\mathcal{O}(\mathcal{G}_2)$ is suggested whose elements are functions from the partially ordered set $\mathcal{C}(\mathbb{C}^2)$ of all unital commutative sub-$\mathrm{C}\!\begin{smallmatrix}{\ast}\\ { } \end{smallmatrix}\!$-algebras $\mathcal{C}$ of $\mathbb{C}^2$ to the poset $\mathcal{L}(\mathbb{C}^2)$. As the result of this, instead of being associated with a single closed subspace of $\mathbb{C}^2$, a proposition corresponds to a family of the subspaces, one family per classical context. The suggested lattice $\mathcal{O}(\mathcal{G}_2)$ is the topology of the quantum phase space $\mathcal{G}_2$, and as such defines a Heyting algebra.\\

\noindent Be this as it may, it should, however, be noted that the motivation for any alternative lattice is somewhat weaker than the motivation for the Hilbert lattice.\\

\noindent Furthermore, while in quantum logic, the assignment of truth-values to the elements of the Hilbert lattice takes the values from a set like $\{0 ,1\}$ which is identified with \textit{false} and \textit{true}, in the intuitionistic quantum logic proposed in \cite{Caspers}, the truth assignment on $\mathcal{O}(\mathcal{G}_2)$ is required to take values from some general and abstract ``truth object'' (like a topos) whose semantical interpretation is not that clear.\\

\section{Supervaluational quantum logic} \label{III} 

\noindent Now, let us analyze the statement that quantum propositions obey PEM.\\

\noindent We will start with the observation made in \cite{Coecke}: Let $A$ and $B$ be the verifiable propositions relating to the quantum-mechanical system. The proposition is called \textit{actual for a particu\/lar realization of the system} (where by particular realization one can understand the quantum state $|\Psi\rangle$ in which the system is prepared) if this proposition has a definite truth value. According to \cite{Coecke}, actuality of $A \vee B$ does not necessarily imply actuality of $A$ or actuality of $B$, i.e., there exists a state (a superposition of states) for which $A \vee B$ is actual, but neither $A$ nor $B$ are actual.\\

\noindent This observation speaks in favor of \textit{supervaluationism}. Let us briefly recall some definitions regarding supervaluationism needed in this paper.\\

\noindent In a word, supervaluationism is a semantics that allows one to apply the tautologies of propositional logic in cases where truth values are undefined \cite{Varzi}.\\

\noindent Supervaluationism retains the classical consequence relation and classical laws whilst admitting \textit{truth-value gaps} (meaning that some propositions have absolutely no truth-value). Accordingly, a disjunction as well as a conjunction may have a definite truth value even when its components do not \cite{Keefe}.\\

\noindent For example, given that the concept of \textit{a heap} lacks sharp boundaries, the proposition ``$N$ grain(s) of wheat is a heap'' \textit{cannot have a truth-value} since no one grain of wheat can be identified as making the difference between being a heap and not being a heap.\\

\noindent However, it is logically true for any number of grains of wheat that it either does or does not make a heap. In other words, the disjunction of the propositions $P$ = ``$N$ grain(s) of wheat is a heap'' and $\neg P$ = ``$N$ grain(s) of wheat is not a heap'' is an instance of the valid schema $P \vee \neg P$ and so, according to supervaluationism, it should be true regardless of whether or not its disjuncts have a truth value; that is, it should be true in all interpretations (in the given example, for any number of grains $N$). As a consequence, supervaluation semantics is no longer truth-functional.\\

\noindent If, in general, something is true in all interpretations, supervaluationism describes it as ``supertrue'', while something false in all interpretations is described as ``superfalse'' \cite{Fine}.\\

\noindent From a mathematical point of view, to interpret quantum propositions in terms of a supervaluationary logic, one must impose upon the closed linear subspaces of the Hilbert space \textit{a structure that allows truth-value gaps}. The collection of invariant-subspace lattices that have no mutual nontrivial members provides a natural candidate for such a structure.\\

\noindent Particularly, to form this ``gappy'' structure, it is enough to strengthen the assumption of the Hilbert lattice $(\mathcal{L}(\mathbb{C}^2),\le)$. That is, the logical predicate of the assumption of the Hilbert lattice\smallskip

\begin{equation}  
   \Phi\left( \mathcal{H}^\prime \right)
   \equiv
   \mathcal{H}^\prime
   \subseteq
   \mathbb{C}^2
   \;\;\;\;  ,
\end{equation}
\smallskip

\noindent i.e., the rule defining the set $\mathcal{L}(\mathbb{C}^2)$ of all the closed linear subspaces $\mathcal{H}^\prime$ of the Hilbert space $\mathbb{C}^2$ using set-builder notation\smallskip

\begin{equation}  
   \mathcal{L}(\mathbb{C}^2)
   =
   \left\{
      \mathcal{H}^\prime
      :\;
      \Phi\left( \mathcal{H}^\prime \right)
   \right\}
   \;\;\;\;  ,
\end{equation}
\smallskip

\noindent should be replaced by a stronger predicate, namely\smallskip

\begin{equation}  
   \Phi\left( \mathcal{H}^\prime, \hat{P}_m^{(R)} \right)
   \equiv
   \Phi\left( \mathcal{H}^\prime \right)
   \;\text{and}\;\,
   \hat{P}_m^{(R)}\!
   :\,
   \mathcal{H}^\prime
   \mapsto
   \mathcal{H}^\prime
   \;\;\;\;  .
\end{equation}
\smallskip

\noindent As per this strengthened rule, each partially ordered set\smallskip

\begin{equation}  
   \mathcal{L}\left( \hat{P}_m^{(R)} \right)
   \equiv
   \left\{
      \mathcal{H}^\prime
      :\;
      \Phi\left( \mathcal{H}^\prime, \hat{P}_m^{(R)} \right)
   \right\}
   \;\;\;\;   
\end{equation}
\smallskip

\noindent can include only those closed linear subspaces $\mathcal{H}^\prime$ that are \textit{invariant under the projection operator $\hat{P}_m^{(R)}$}. That is, the image of every vector $|\Psi\rangle$ in those $\mathcal{H}^\prime$ under $\hat{P}_m^{(R)}$ remains within $\mathcal{H}^\prime$ which can be denoted as\smallskip

\begin{equation}  
   \hat{P}_m^{(R)} \mathcal{H}^\prime
   \equiv
   \left\{
      |\Psi\rangle
      \in
      \mathcal{H}^\prime
      \!:\,
      \hat{P}_m^{(R)} |\Psi\rangle
   \right\}
   \subset
   \mathcal{H}^\prime     
   \;\;\;\;  .
\end{equation}
\smallskip

\noindent The elements of every set $\mathcal{L}(\hat{P}_m^{(R)})$, explicitly,\smallskip

\begin{equation}  
   \mathcal{L} \left( \hat{P}_m^{(R)} \right)
   =
   \left\{
      \mathrm{ran}(\hat{0})
      ,\,
      \mathrm{ran}(\hat{P}_m^{(R)})
      ,\,
      \mathrm{ran}(\neg \hat{P}_m^{(R)})
      ,\,
      \mathrm{ran}(\hat{1})\!
   \right\}
   \;\;\;\;  ,
\end{equation}
\smallskip

\noindent form \textit{the invariant-subspace lattice $(\mathcal{L}(\hat{P}_m^{(R)}),\le)$}, a complete complemented distributive lattice (a Boolean algebra) \cite{Radjavi}. As it is obvious, each set $\mathcal{L}(\hat{P}_m^{(R)})$ only contains the closed linear subspaces belonging to the mutually commutable projection operators.\\

\noindent From here one can conclude that the nontrivial closed linear subspace $\mathrm{ran}(\hat{P}_m^{(R)})$ can only belong to the poset $\mathcal{L}(\hat{P}_m^{(R)})$ while $\mathrm{ran}(\hat{P}_n^{(Q)})$, where $Q \neq R$, – only to $\mathcal{L}(\hat{P}_n^{(Q)})$. Being elements of different lattices, these nontrivial closed linear subspaces cannot meet each other. In symbols,\smallskip

\begin{equation}  
   Q \neq R
   ,\,
   \mathrm{ran}(\hat{P}_n^{(Q)})
   \in
   \mathcal{L}(\hat{P}_n^{(Q)})
   ,\,
   \mathrm{ran}(\hat{P}_m^{(R)})
   \in
   \mathcal{L}(\hat{P}_m^{(R)})
   \,
   \implies
   \,
   \mathrm{ran}(\hat{P}_n^{(Q)})
   \;\cancel{\;\wedge\;}\;
   \mathrm{ran}(\hat{P}_m^{(R)})
   \;\;\;\;  ,
\end{equation}
\smallskip

\noindent where the cancelation of $\wedge$ indicates that this operation cannot be defined.\\

\noindent The nonexistence of the meet operation for pairs of the ranges that do not lie in a common invariant-subspace lattice corresponds to truth-value gaps in the supervaluational logic.\\

\noindent To be sure, consider the proposition $P_m^{(R)}$ and its negation $\neg P_m^{(R)}$. If the realization of the qubit is given by the state $|\Psi_n^{(Q)}\rangle$, they are:\smallskip

\begin{equation} \label{P1} 
   P_m^{(R)}\!
   =
   \mathrm{Prop}\!
   \left(
      |\Psi_n^{(Q)}\rangle
      \in
      \mathrm{ran}(\hat{P}_n^{(Q)})
      \;\cancel{\;\wedge\;}\;
      \mathrm{ran}(\hat{P}_m^{(R)})
   \right)
   \;\;\;\;  ,
\end{equation}

\begin{equation} \label{P2} 
   \neg P_m^{(R)}\!
   =
   \mathrm{Prop}\!
   \left(
      |\Psi_n^{(Q)}\rangle
      \in
      \mathrm{ran}(\hat{P}_n^{(Q)})
      \;\cancel{\;\wedge\;}\;
      \mathrm{ran}(\hat{P}_{m+(-1)^{m-1}}^{(R)})
   \right)
   \;\;\;\;  .
\end{equation}
\smallskip

\noindent As the operation $\cancel{\;\wedge\;}$ is undefined, the proposition $P_m^{(R)}$ and its negation $\neg P_m^{(R)}$ cannot have a truth value in the state $|\Psi_n^{(Q)}\rangle$ (to borrow the terminology from the paper \cite{Coecke}, \textit{they are not actual in this state}), that is,\smallskip

\begin{equation}  
   b\left(
      P_m^{(R)}\!
   \right)
   =
   b\bigg(
      \mathrm{Prop}\!
      \left(
         |\Psi_n^{(Q)}\rangle
         \in
         \mathrm{ran}(\hat{P}_n^{(Q)})
         \;\cancel{\;\wedge\;}\;
         \mathrm{ran}(\hat{P}_m^{(R)})
      \right)\!\!
   \bigg)
   =
   \frac{0}{0}
   \;\;\;\;  ,
\end{equation}

\begin{equation}  
   b\left(
      \neg P_m^{(R)}\!
   \right)
   =
   b\bigg(
      \mathrm{Prop}\!
      \left(
         |\Psi_n^{(Q)}\rangle
         \in
         \mathrm{ran}(\hat{P}_n^{(Q)})
         \;\cancel{\;\wedge\;}\;
         \mathrm{ran}(\hat{P}_{m+(-1)^{m-1}}^{(R)})
      \right)\!\!
   \bigg)
   =
   \frac{0}{0}
   \;\;\;\;  ,
\end{equation}
\smallskip

\noindent where $b$ stands for \textit{the bivalent valuation relation}, i.e., the function from the set of propositions into the set $\{0,1\}$ of bivalent truth values, and $\frac{0}{0}$ denotes an indeterminate value.\\

\noindent In contrast to this, the propositions\smallskip

\begin{equation} \label{TAUT} 
   \mathrm{Prop}
   \bigg(
      |\Psi_n^{(Q)}\rangle
      \in
      \mathrm{ran}(\hat{P}_n^{(Q)})
      \wedge
      \left(
         \mathrm{ran}(\hat{P}_m^{(R)})
         \vee
         \mathrm{ran}(\neg \hat{P}_m^{(R)})
      \right)\!\!
   \bigg)
   \;\;\;\;   \text{and}
\end{equation}

\begin{equation} \label{CONT} 
   \mathrm{Prop}
   \bigg(
      |\Psi_n^{(Q)}\rangle
      \in
      \mathrm{ran}(\hat{P}_n^{(Q)})
      \wedge
      \left(
         \mathrm{ran}(\hat{P}_m^{(R)})
         \wedge
         \mathrm{ran}(\neg \hat{P}_m^{(R)})
      \right)\!\!
   \bigg)
   \qquad \quad
\end{equation}
\smallskip

\noindent have a defined truth-value because\smallskip

\begin{equation}  
   \mathrm{ran}(\hat{P}_m^{(R)})
   \vee
   \mathrm{ran}(\neg \hat{P}_m^{(R)})
   =
   \mathbb{C}^2
   \;\;\;\;   \text{and}
\end{equation}

\begin{equation}  
   \mathrm{ran}(\hat{P}_m^{(R)})
   \wedge
   \mathrm{ran}(\neg \hat{P}_m^{(R)})
   =
   \{0\}
   \qquad \quad
\end{equation}
\smallskip

\noindent are the trivial elements of every invariant-subspace lattice and, hence, can meet the subspace $\mathrm{ran}(\hat{P}_n^{(Q)})$. In symbols,\smallskip

\begin{equation}  
   \left\{
      \{0\}
      ,\,
      \mathrm{ran}(\hat{P}_n^{(Q)})
   \right\}
   \subset
   \mathcal{L}(\hat{P}_n^{(Q)})
   \,
   \implies
   \,
   \{0\}
   \wedge
   \mathrm{ran}(\hat{P}_n^{(Q)})
   \in
   \mathcal{L}(\hat{P}_n^{(Q)})
   \;\;\;\;  ,
\end{equation}

\begin{equation}  
   \left\{
      \mathrm{ran}(\hat{P}_n^{(Q)})
      ,\,
      \mathbb{C}^2
   \right\}
   \subset
   \mathcal{L}(\hat{P}_n^{(Q)})
   \,
   \implies
   \,
   \mathrm{ran}(\hat{P}_n^{(Q)})
   \wedge
   \mathbb{C}^2
   \in
   \mathcal{L}(\hat{P}_n^{(Q)})
   \;\;\;\;  .
\end{equation}
\smallskip

\noindent Given\smallskip

\begin{equation}  
   b\bigg(
      \mathrm{Prop}\!
      \left(
         |\Psi_n^{(Q)}\rangle
         \in
         \mathrm{ran}(\hat{P}_n^{(Q)})
      \right)\!\!
   \bigg)
   =
   1
   \;\;\;\;  ,
\end{equation}

\begin{equation}  
   b\bigg(
      \mathrm{Prop}\!
      \left(
         |\Psi_n^{(Q)}\rangle
         \in
         \{0\}
      \right)\!\!
   \bigg)
   =
   0
   \;\;\;\;  ,
\end{equation}
\smallskip

\noindent the propositions (\ref{TAUT}) and (\ref{CONT}) are, in terms of supervaluationism, \textit{supertrue} and \textit{superfalse}, correspondingly.\\

\noindent On the other hand, seeing as the ranges $\mathrm{ran}(\hat{P}_m^{(R)})$ and $\mathrm{ran}(\neg \hat{P}_m^{(R)})$ are members of the Boolean lattice $(\mathcal{L}(\hat{P}_m^{(R)}),\le)$, one can maintain a classical logical interpretation of the meet and join of these subspaces and consequently present\smallskip

\begin{equation}  
   P_m^{(R)}\!
   \vee
   \neg P_m^{(R)}
   =
   \mathrm{Prop}
   \left(
      |\Psi_n^{(Q)}\rangle
      \in
      \mathrm{ran}(\hat{P}_n^{(Q)})
      \wedge
      \mathbb{C}^2
   \right)
   =
   \mathrm{Prop}
   \left(
      |\Psi_n^{(Q)}\rangle
      \in
      \mathrm{ran}(\hat{P}_n^{(Q)})
   \right)
   \;\;\;\;  ,
\end{equation}

\begin{equation}  
   P_m^{(R)}\!
   \wedge
   \neg P_m^{(R)}
   =
   \mathrm{Prop}
   \left(
      |\Psi_n^{(Q)}\rangle
      \in
      \mathrm{ran}(\hat{P}_n^{(Q)})
      \wedge
      \{0\}
   \right)
   =
   \mathrm{Prop}
   \left(
      |\Psi_n^{(Q)}\rangle
      \in
      \{0\}
   \right)
   \;\;\;\;  .
\end{equation}
\smallskip

\noindent Therefore, even though the proposition $P_m^{(R)}$ and its negation $\neg P_m^{(R)}$ have no truth value in the state $|\Psi_n^{(Q)}\rangle$, their disjunction and conjunction are true and false, respectively, under any possible realization of the qubit, namely, $b(P_m^{(R)}\! \vee \neg P_m^{(R)}) = 1$ and $b(P_m^{(R)}\! \wedge \neg P_m^{(R)}) = 0$.\\

\noindent In this sense, the supervaluational semantics of quantum propositions does not violate the classical principles of excluded middle and non-contradiction (according to which a proposition and its negation cannot be both true \cite{Beziau}).\\

\section{Many-valued quantum logic}  

\noindent One may say that ``gappy'' propositions like (\ref{P1}) and (\ref{P2}) have no truth value for the reason that they do not belong to the domain of two-valued logic. Therefore, one may assign to ``gappy'' propositions a new – i.e., different from \textit{true} and \textit{false} – truth-value (called, for example, ``undetermined'') and assume that the image of this new truth-value under the valuation relation lies between 0 and 1. By doing so, one can construct a many-valued semantics of quantum propositions which defines the same logic as the supervaluational semantics does.\\

\noindent For example, in the infinite-valued semantics of quantum propositions proposed in a series of works \cite{Pykacz94, Pykacz95, Pykacz10, Pykacz11, Pykacz15}, the valuation $v$ (i.e., the function from the set of propositions into the interval $[0,1]$ of the infinite-valued truth degrees) is set forth by\smallskip

\begin{equation}  
   v\left(
      P_n^{(Q)}\!
   \right)
   =
   \langle \Psi| \hat{P}_n^{(Q)}\! |\Psi\rangle
   \in
   [0,1]
   \;\;\;\;  .
\end{equation}
\smallskip

\noindent As it can be readily seen from here, if the realization of the qubit is given by the state $|\Psi_m^{(R)}\rangle$, then the truth degree of the proposition $P_1^{(R)}$ must be $\langle \Psi_m^{(R)}| \hat{P}_1^{(R)}\! |\Psi_m^{(R)}\rangle$, which is 1 or 0.\\ 

\noindent By contrast, in the case where the realization of the qubit is given by the superposition of the states $|\Psi_1^{(R)}\rangle$ and $|\Psi_2^{(R)}\rangle$, the truth degree of the proposition $P_m^{(R)}$ and its negation $\neg P_m^{(R)}$ are\smallskip

\begin{equation} \label{VP1} 
   v\left(
      P_m^{(R)}
   \right)
   =
   \langle \Psi_n^{(Q)}| \hat{P}_m^{(R)}\! |\Psi_n^{(Q)}\rangle
   \in
   (0,1)
   \;\;\;\;  ,
\end{equation}

\begin{equation} \label{VP2} 
   v\!\left(
      \neg P_m^{(R)}
   \right)
   =
   \langle \Psi_n^{(Q)}| \neg \hat{P}_m^{(R)}\! |\Psi_n^{(Q)}\rangle
   =
   \left(
      1
      -
      \langle \Psi_n^{(Q)}| \hat{P}_m^{(R)}\! |\Psi_n^{(Q)}\rangle
   \right)
   \in\!
   (0,1)
   \;\;\;\;  .
\end{equation}
\smallskip

\noindent In this way, what is regarded as truth-value gaps in the supervaluational semantics is filled out with the truth degrees lying between 0 and 1 in the many-valued semantics.\\

\noindent However, the many-valued semantics exhibits a problem pertaining to \textit{the interpretation of the truth degrees}. To illustrate this problem, suppose that the gaps $b(P_1^{(R)}) = \frac{0}{0}$ and $b(\neg P_1^{(R)}) = \frac{0}{0}$ are filled with the truth degrees in a manner that $v(P_1^{(R)}) \neq v(\neg P_1^{(R)})$, which, according to (\ref{VP1}) and (\ref{VP2}), entails $\langle \Psi_n^{(Q)}| \hat{P}_1^{(R)}\! |\Psi_n^{(Q)}\rangle \neq \frac{1}{2}$. The question is, what does such a difference mean?\\

\noindent The difficulty with this question is that there does not exist a standard interpretation of the truth degrees and, therefore, how the difference $\langle \Psi_n^{(Q)}| \hat{P}_1^{(R)}\! |\Psi_n^{(Q)}\rangle \neq \frac{1}{2}$ is to be understood depends on the chosen interpretation of quantum mechanics.\\

\noindent Thus, in Quantum Bayesianism, i.e., the Bayesian approach to quantum mechanics \cite{Caves, Fuchs}, the said difference represents the inequality in the degrees of belief of an agent regarding the proposition  $P_1^{(R)}$ and its negation. Whereas, under the Copenhagen interpretation, the same difference may be understood as the contrast in the degrees to which the qubit possesses and does not possess the property $P_1^{(R)}$ before its verification.\\

\section{Counterfactual definiteness of quantum logic}  

\noindent From what is argued in the Section \ref{III}, it follows that the failure of PEM demonstrated in (\ref{NEG1}) and (\ref{NEG2}) is caused by the ``gapless'' structure of the Hilbert lattice $(\mathcal{L}(\mathbb{C}^2),\le)$.\\

\noindent Given $\Sigma$, the collection of all the nontrivial projection operators on $\mathbb{C}^2$,  namely,\smallskip

\begin{equation}  
   \Sigma
   =
   \left\{
      \Sigma^{(R)})
   \right\}_{R=1}^3
   =
   \left\{
      \hat{P}_1^{(R)}
      ,\,
      \hat{P}_2^{(R)}
   \right\}_{R=1}^3
   \;\;\;\;  ,
\end{equation}
\smallskip 

\noindent the assumption of the Hilbert lattice is formally equivalent to the statement that for the collection of $\mathcal{L}(\hat{P}_m^{(R)})$, there exists a set-theoretic union $\mathcal{L}(\mathbb{C}^2)$, i.e.,\smallskip

\begin{equation} \label{STAT} 
   \mathcal{L}(\mathbb{C}^2)
   =
   \bigcup_{\hat{P}_m^{(R)} \in \Sigma}
      \mathcal{L}(\hat{P}_m^{(R)})
   \;\;\;\;  ,
\end{equation}
\smallskip 

\noindent or, explicitly,\smallskip

\begin{equation}  
   \left\{
      \{0\}
      ,
      \mathrm{ran}(\hat{P}_{1}^{(1)\!})
      ,
      \mathrm{ran}(\hat{P}_{2}^{(1)\!})
      ,
      \dots
      ,
      \mathrm{ran}(\hat{P}_{2}^{(3)\!})
      ,
      \mathbb{C}^2\!
   \right\}
   =
   \!\!\!\!
   \bigcup_{\hat{P}_m^{(R)} \in \Sigma}
   \!\!\!\!\!
   \left\{
      \{0\}
      ,
      \mathrm{ran}(\hat{P}_m^{(R)}\!)
      ,
      \mathrm{ran}(\hat{P}_{m+(-1)^{m-1}}^{(R)}\!)
      ,
      \mathbb{C}^2\!
   \right\}
   \;\;\;\;  .
\end{equation}
\smallskip

\noindent This statement brings on the meet operation on any pairs of the closed subspaces of $\mathbb{C}^2$, i.e.,\smallskip

\begin{equation} \label{MEET} 
   \left\{
      \mathrm{ran}(\hat{P}_n^{(Q)})
      ,\,
      \mathrm{ran}(\hat{P}_m^{(R)})
   \right\}
   \subset
   \!\!\!\!
   \bigcup_{\hat{P}_m^{(R)} \in \Sigma}
   \!\!\!\!\!
      \mathcal{L}(\hat{P}_m^{(R)})
   \,
   \implies
   \,
   \mathrm{ran}(\hat{P}_n^{(Q)})
   \wedge
   \mathrm{ran}(\hat{P}_m^{(R)})
   \in
   \!\!\!\!
   \bigcup_{\hat{P}_m^{(R)} \in \Sigma}
   \!\!\!\!\!
      \mathcal{L}(\hat{P}_m^{(R)})
   \;\;\;\;  ,
\end{equation}
\smallskip

\noindent thus implying that a proposition and its negation may be both false.\\

\noindent Semantically, though, the statement (\ref{STAT}) is consistent with the assumption of \textit{counterfactual definiteness}.\\

\noindent Recall that counterfactual definiteness is the ability to speak meaningfully of the definiteness of the results of measurements that have not been performed \cite{Hess}. Equally, this term can be used to imply the ability to assign a definite truth value to a proposition that has not yet been verified.\\

\noindent Along the lines of counterfactual definiteness, the existence of the meet operation on the column spaces of the incommutable projection operators $\hat{P}_n^{(Q)}$ and $\hat{P}_m^{(R)}$ stated in (\ref{MEET}) means that the proposition like $\mathrm{Prop}\big(|\Psi\rangle \!\in\! \mathrm{ran}(\hat{P}_n^{(Q)}) \wedge \mathrm{ran}(\hat{P}_1^{(R)})\big)$ asserting that ``Had the verification of $P_n^{(Q)} = \mathrm{Prop}\big(|\Psi\rangle \in \mathrm{ran}(\hat{P}_n^{(Q)})\big)$ been performed, the truth value of $P_1^{(R)}$ would have been obtained'' always has a definite truth value even though the simultaneous verifications of $P_n^{(Q)}$ and $P_1^{(R)}$ were not carried due to the incommutability of $\hat{P}_n^{(Q)}$ and $\hat{P}_m^{(R)}$.\\

\noindent In this way, the ``gappy'' lattice structure of the supervaluationary logic of quantum propositions rejects counterfactual definiteness inasmuch as a truth value of the proposition $\mathrm{Prop}\big(|\Psi\rangle \!\in\! \mathrm{ran}(\hat{P}_n^{(Q)}) \,\cancel{\wedge}\, \mathrm{ran}(\hat{P}_1^{(R)})\big)$ cannot be defined.\\

\section{Concluding remarks}  

\noindent Being omnipresent in classical logic, the principle of excluded middle plays an important role in the issue of \textit{macroscopic realism} \cite{Reid}.\\

\noindent In terms of Schrodinger's cat gedanken experiment \cite{Schrodinger} (where the premise is that the macroscopically distinguishable states ``dead'' and ``alive'' are the quantum states of the ``cat''), given that after the verification of its status, the cat can be only found in one of the two quantum states – either dead or alive – macroscopic realism asserts that \textit{the cat is always is in one of these states}, even before the verification. As a consequence, the proposition $P$ = ``the cat is dead'' and its negation $\neg P$ = ``the cat is alive'' cannot be both false not only after the verification but also prior to the verification. Hence, macroscopic realism implies that the quantum propositions obey PEM.\\

\noindent Typically, to negate macroscopic realism, the rejection of PEM is considered and, consequently, an intuitionistic approach to quantum logic is sought (see, e.g., \cite{Caspers}). However, as it has been shown in this paper, supervaluational quantum logic refutes macroscopic realism as well.\\

\noindent To be sure, according to the supervaluationist account of Schrodinger's cat gedanken experiment, when the cat is in the superposition state, the disjunction $P\vee\neg P$ and conjunction $P\wedge\neg P$ are true and false, correspondingly, but neither $P$ nor $\neg P$ is actual. That is, the statement ``Out of two possible states, dead and alive, the cat is in one or the other but not in both'' is true despite the fact the statement ``The cat is in one of these states'' has absolutely no truth-value before the verification.\\

\bibliographystyle{References}
\bibliography{PEM_Ref}

\end{document}